\documentclass{llncs}  

\usepackage{amssymb}
\usepackage{xspace}
\usepackage[english]{babel}
\usepackage{amsmath}
\usepackage{latexsym}
\usepackage{textcomp}
\usepackage{epsfig}
\usepackage{subfigure}
\usepackage{color}
\usepackage{wrapfig}

\usepackage[labelfont=bf,size=small]{caption}
\usepackage[plainpages=false]{hyperref}
\usepackage{calc}
\usepackage[textsize=footnotesize]{todonotes}
\usepackage{lineno}

\newcommand{\remove}[1]{}

\newcommand{\CCT}{CCR\xspace}
\newcommand{\UCCT}{UCCR\xspace}
\newcommand{\algo}{\texttt{UCCRDrawer}\xspace}
\newcommand{\balanced}{balanced\xspace}
\newcommand{\spaced}{well-spaced\xspace}
\newcommand{\saferegion}{safe-region\xspace}
\newcommand{\upslopeset}{slope-set\xspace}
\newcommand{\tsp}{SP}
\newcommand{\ttsp}{\tsp{-digraph}\xspace}
\newcommand{\ttspg}{\tsp{-graph}\xspace}

\title{1-bend Upward Planar Drawings of SP-digraphs\thanks{Research supported in part by the MIUR project AMANDA ``Algorithmics for MAssive and Networked DAta'', prot. 2012C4E3KT\_001.}}
\author{Emilio Di Giacomo, Giuseppe Liotta, Fabrizio Montecchiani}

\institute{Dip. di Ingegneria,
Universit\`a degli Studi di Perugia\\ \email{\{emilio.digiacomo,giuseppe.liotta,fabrizio.montecchiani\}@unipg.it} }

\begin{document}

\maketitle


\begin{abstract}
It is proved that every series-parallel digraph whose maximum vertex-degree is $\Delta$ admits an upward planar drawing with at most one bend per edge such that each edge segment has one of $\Delta$ distinct slopes. This is shown to be worst-case optimal in terms of the number of slopes. 
Furthermore, our construction gives rise to drawings with optimal angular resolution $\frac{\pi}{\Delta}$.
A variant of the proof technique is used to show that (non-directed)  reduced series-parallel graphs and flat series-parallel graphs have a (non-upward) one-bend planar drawing with $\lceil\frac{\Delta}{2}\rceil$ distinct slopes if biconnected, and with $\lceil\frac{\Delta}{2}\rceil+1$ distinct slopes if connected.
\end{abstract}

\section{Introduction}

The \emph{$k$-bend planar slope number} of a family of planar graphs with maximum vertex-degree $\Delta$ is the minimum number of distinct slopes used for the edges when computing a crossing-free drawing with at most $k > 0$ bends per edge of any graph in the family. For example, if $\Delta = 4$, a classic result is that every planar graph has a crossing-free drawing such that every edge segment is either horizontal or vertical and each edge has at most two bends (see, e.g.,~\cite{DBLP:journals/comgeo/BiedlK98}). Clearly, this is an optimal bound on the number of slopes. This result has been extended to values of $\Delta$ larger than four by Keszegh et al.~\cite{DBLP:journals/siamdm/KeszeghPP13}, who prove that $\lceil\frac{\Delta}{2}\rceil$ slopes suffice to construct a planar drawing with at most two bends per edge for any planar graph. However, if additional geometric constraints are imposed on the crossing-free drawing, only a few tight bounds on the planar slope number are known. For example, if one requires that the edges cannot have bends,  the best known upper bound on the planar slope number is $O(c^{\Delta})$ (for a constant $c > 1$) while a general lower bound of just $3\Delta -6$ has been proved~\cite{DBLP:journals/siamdm/KeszeghPP13}. Tight bounds are only known for outerplanar graphs~\cite{DBLP:journals/comgeo/KnauerMW14} and subcubic planar graphs~\cite{DBLP:conf/latin/GiacomoLM14}, while the gap between upper and lower bound has been reduced for planar graphs with treewidth two~\cite{DBLP:conf/gd/LenhartLMN13} or three~\cite{DBLP:journals/jgaa/GiacomoLM15,DBLP:journals/gc/JelinekJKLTV13}. If one bend per edge is allowed, Keszegh et al.~\cite{DBLP:journals/siamdm/KeszeghPP13} show an upper bound of $2 \Delta$ and a lower bound of $\frac{3}{4}(\Delta - 1)$ on the planar slope number of the planar graphs with maximum vertex-degree $\Delta$. In a recent paper, Knauer and Walczak~\cite{DBLP:conf/latin/KnauerW16} improve the upper bound to $\frac{3}{2}(\Delta - 1)$; in the same paper, it is also proved that a tight bound of $\lceil\frac{\Delta}{2}\rceil$ can be achieved for the outerplanar graphs.

In this paper we focus on the 1-bend planar slope number of directed graphs with the additional requirement that the computed drawing be  \emph{upward}, i.e., each edge is drawn as a curve monotonically increasing in the $y$-direction.  
We recall that upward drawings are a classic research topic in graph drawing, see, e.g.,~\cite{DBLP:journals/siamcomp/BertolazziBMT98,DBLP:journals/comgeo/BinucciDG08,DBLP:journals/jgaa/Didimo06,DBLP:journals/siamdm/DidimoGL09,DBLP:journals/siamcomp/GargT01} for a limited list of references. Also, upward drawings of ordered sets with no bends and few slopes have been studied by Czyzowicz~\cite{DBLP:journals/jct/Czyzowicz91,DBLP:journals/dm/CzyzowiczPR90}. 
We show that every series-parallel digraph (\ttsp for short) $G$ whose maximum vertex-degree is $\Delta$ has \emph{1-bend upward planar slope number} $\Delta$. That is, $G$ admits an upward planar drawing with at most one bend per edge where at most $\Delta$ distinct slopes are used for the edges. This is shown to be worst-case optimal in terms of the number of slopes. 
An implication of this result is that the general $\frac{3}{2}(\Delta - 1)$ upper bound for the (undirected) 1-bend planar slope number~\cite{DBLP:conf/latin/KnauerW16} can be lowered to $\Delta$ when the graph is series-parallel. We then extend our drawing technique to undirected graphs and hence look at non-upward drawings. We show a tight bound of $\lceil\frac{\Delta}{2}\rceil$ for the 1-bend planar slope number of biconnected reduced {\ttspg}s and biconnected flat {\ttspg}s (see Section~\ref{se:preliminaries} for  definitions). The biconnectivity requirement can be dropped at the expenses of one more slope. 
To prove the above results, we construct a suitable contact representation $\gamma$ of an \ttsp where each vertex is represented as a cross, i.e. a horizontal segment intersected by a vertical segment (Section~\ref{se:ccr}); then, we transform $\gamma$ into a 1-bend upward planar drawing $\Gamma$ optimizing the number of slopes used in such transformation (Section~\ref{se:upward}). 
Our algorithm runs in linear time and gives rise to drawings with angular resolution at least $\frac{\pi}{\Delta}$, which is worst-case optimal. 
Some proofs and technicalities can be found in the appendix.

\section{Preliminaries}\label{se:preliminaries}
\begin{figure}[t]
\centering
\subfigure[]{\includegraphics[scale=0.8,page=2]{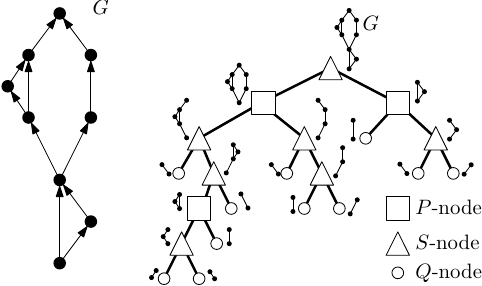}\label{fi:ttsp}}\hfil
\subfigure[]{\includegraphics[scale=0.9,page=3]{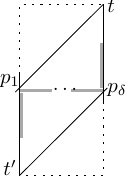}\label{fi:safe-rectangle}}
\caption{
(a) An {\ttsp} $G$ and its decomposition tree. 
(b) The \saferegion (dotted) of a cross.
}
\end{figure}

A \emph{series-parallel digraph} (\emph{\ttsp} for short)~\cite{DBLP:books/ph/BattistaETT99} is a simple planar digraph that has one source and one sink, called \emph{poles}, and it is recursively defined as follows. A single edge is an {\ttsp}. The digraph obtained by identifying the sources and the sinks of two {\ttsp{s}} is an {\ttsp} (\emph{parallel composition}). The digraph obtained by identifying the sink of one {\ttsp} with the source of a second {\ttsp} is an {\ttsp} (\emph{series composition}). A \emph{reduced} \ttsp is an \ttsp with no transitive edges. 
An {\ttsp} $G$ is associated with a binary tree $T$, called the \emph{decomposition tree} of $G$.
The nodes of $T$ are of three types, \emph{$Q$-nodes}, \emph{$S$-nodes}, and \emph{$P$-nodes}, representing single edges, series compositions, and parallel compositions, respectively.
An example is shown in Fig.~\ref{fi:ttsp}. 
The decomposition tree of $G$ has $O(n)$ nodes and can be constructed in $O(n)$ time~\cite{DBLP:books/ph/BattistaETT99}. 
An \ttsp is \emph{flat} if its decomposition tree does not contain two $P$-nodes that share only one pole and that are not in a series composition (see, e.g.,~\cite{DBLP:conf/gd/Giacomo03}). The underlying undirected graph of an {\ttsp} is called an \emph{\ttspg}, and the definitions of reduced and flat \ttsp{s} translate to it.

The \emph{slope} $s$ of a line $\ell$ is the angle that a horizontal line needs to be rotated counter-clockwise in order to make it overlap with $\ell$. The slope of a segment is the slope of its supporting line. We denote by $\mathcal S_k$ the set of slopes: $s_i = \frac{\pi}{2} + i \frac{\pi}{k}$ ($i=0,\dots,k-1$). 
Note that $\mathcal S_k$ contains the slope $\frac{\pi}{2}$ for any value of $k$. 
Also, any polyline drawing using only slopes in $\mathcal S_k$ has angular resolution (i.e. the minimum angle between any two consecutive edges around a vertex) at least $\frac{\pi}{k}$.

\section{Cross Contact Representations}\label{se:ccr}

\smallskip\noindent{\bf \em Basic definitions.} A \emph{cross} consists of one horizontal and one vertical segment that share an interior point, called \emph{center} of the cross.  A cross is \emph{degenerate} if either its horizontal or its vertical segment has zero length. The center of a degenerate cross is its midpoint. A point $p$ of a cross $c$ is an \emph{end-point} (\emph{interior point}) of $c$ if it is an end-point (interior point) of the horizontal or vertical segment of $c$. Two crosses $c_1$ and $c_2$ \emph{touch} if they share a point $p$, called \emph{contact}, such that $p$ is an end-point of the vertical (horizontal) segment of $c_1$ and an interior point of the horizontal (vertical) segment of $c_2$. A \emph{cross-contact representation (\CCT)} of a graph $G$ is a drawing $\gamma$ such that: $(i)$ Every vertex $v$ of $G$ is represented by a  cross $c(v)$; $(ii)$ All intersections of crosses are contacts; and $(iii)$ Two crosses $c(u)$ and $c(v)$ touch if and only if the edge $(u,v)$ is in $G$. 

We now consider {\CCT}s of digraphs, and define properties that will be useful to transform a \CCT into a $1$-bend upward planar drawing with few slopes and good angular resolution. Let $\gamma$  be a \CCT of a digraph $G$ with maximum vertex-degree $\Delta$. Let  $(u,v)$ be an edge of $G$ oriented from $u$ to $v$. Let $p$ be the contact between $c(u)$ and $c(v)$. The point $p$ is an \emph{upward contact} if the following two conditions hold: (a) $p$ is an end-point of the vertical segment of one of the two crosses and an interior point of the other cross, and (b) the center of $c(v)$ is above the center of $c(u)$. A \CCT of a digraph $G$ such that all its contacts are upward is an \emph{upward \CCT (\UCCT)}. An \UCCT $\gamma$ is \emph{\balanced} if for every non-degenerate cross $c(u)$ of $\gamma$, we have that $|n_l(u) - n_r(u)| \leq 1$, where $n_l(u)$ ($n_r(u)$) is the number of contacts to the left (right) of the center of $c(u)$. Let  $\{p_1,p_2,\dots,p_\delta\}$ be the  $\delta \geq 0$ contacts along the horizontal segment of $c(u)$, in this order from the leftmost one  ($p_1$) to the rightmost one ($p_\delta$). Let $t$ be the intersection point between the vertical line passing through $p_\delta$ and the line with slope $\frac{\pi}{2}-\frac{\pi}{\Delta}$ and passing through $p_1$. Similarly, let $t'$ be the intersection point between the vertical line passing through $p_1$ and the line with slope $\frac{\pi}{2}-\frac{\pi}{\Delta}$ and passing through $p_\delta$. The \emph{\saferegion} of $c(u)$ is the rectangle having $t$ and $t'$ as the top-right and bottom-left corner, respectively. See Fig.~\ref{fi:safe-rectangle} for an illustration. If $\delta=1$, the \saferegion degenerates to a point, while it is not defined when $\delta=0$. An \UCCT $\gamma$ is \emph{\spaced} if no two {\saferegion}s intersect each other.

\smallskip\noindent{\bf \em Drawing construction.} We describe a linear-time algorithm, \algo, that takes as input a reduced \ttsp $G$, and computes an \UCCT $\gamma$ of $G$ that is \balanced and \spaced. The algorithm computes $\gamma$ through a bottom-up visit of the decomposition tree $T$ of $G$. For each node $\mu$ of $T$, it computes an \UCCT $\gamma_\mu$ of the graph $G_\mu$ associated with $\mu$ satisfying the following properties: 
{\bf P1.} $\gamma_\mu$ is \balanced; 
{\bf P2.} $\gamma_\mu$ is \spaced; 
{\bf P3.} Let $s_\mu$ and $t_\mu$ be the two poles of $G_\mu$. If $\mu$ is not a $Q$-node, then both $c(s_\mu)$ and $c(t_\mu)$ are degenerate, with $c(s_\mu)$ at the bottom side of a rectangle $R_\mu$ that contains $\gamma_\mu$, and $c(t_\mu)$ at the top side of $R_\mu$.

\begin{figure}[tbp]
\centering
\subfigure[$Q$ (Type A)]{\includegraphics[scale=1,page=1]{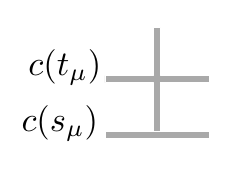}\label{fi:qnode-A}}\hfil
\subfigure[$Q$ (Type B)]{\includegraphics[scale=1,page=2]{figures/qnodes}\label{fi:qnode-B}}\hfil
\subfigure[$S$ (Case 1)]{\includegraphics[scale=1,page=3]{figures/qnodes}\label{fi:snode-1}}\\
\subfigure[$S$ (Case 2)]{\includegraphics[scale=1,page=4]{figures/qnodes}\label{fi:snode-2}}\hfil
\subfigure[$S$ (Case 3)]{\includegraphics[scale=1,page=5]{figures/qnodes}\label{fi:snode-3}}\hfil
\subfigure[$P$]{\includegraphics[scale=1,page=6]{figures/qnodes}\label{fi:pnode}}
\caption{
Illustration for \algo. The {\saferegion}s are dotted (and not in scale).}
\end{figure}


For each leaf node $\mu$ (which is a $Q$-node) the associated graph $G_{\mu}$ consists of a single edge $(s_\mu,t_\mu)$. We define two possible types of \UCCT, $\gamma_\mu^A$ (type A) and $\gamma_\mu^B$ (type B), of $G_\mu$, which are shown in Figs.~\ref{fi:qnode-A} and~\ref{fi:qnode-B}, respectively. Properties {\bf P1 -- P2} trivially hold in this case,  while property {\bf P3} does not apply. 

For each non-leaf node $\mu$ of $T$, \algo computes the \UCCT $\gamma_\mu$ by suitably combining the (already) computed {\UCCT}s $\gamma_{\nu_1}$ and $\gamma_{\nu_2}$ of the two graphs associated with the children $\nu_1$ and $\nu_2$ of $\mu$. 
If $\mu$ is an $S$-node of $T$, we distinguish between the following cases, where $t_{\nu_1} = s_{\nu_2}$ is the pole shared by $\nu_1$ and $\nu_2$. 

\noindent \textbf{Case 1.} Both $\nu_1$ and $\nu_2$ are $Q$-nodes. Then an \UCCT of $G_\mu$ is computed by combining $\gamma_{\nu_1}^A$ and $\gamma_{\nu_2}^B$ as in Fig.~\ref{fi:snode-1}. Properties {\bf P1 -- P3} trivially hold.

\noindent \textbf{Case 2.} $\nu_1$ is a $Q$-node, while $\nu_2$  is not (the case when $\nu_2$ is a $Q$-node and $\nu_1$ is not is symmetric). We combine the drawing $\gamma_{\nu_1}^A$ of $G_{\nu_1}$ and the drawing $\gamma_{\nu_2}$ of $G_{\nu_2}$ as in Fig.~\ref{fi:snode-2}. Notice that to combine the two drawings we may need to scale one of them so that their widths are the same. To ensure {\bf P1}, we move the vertical segment  of $c(t_{\nu_1}) = c(s_{\nu_2})$ so that $|n_l(t_{\nu_1}) - n_r(t_{\nu_1})| \leq 1$. We may also need to shorten its upper part in order to avoid crossings with other segments, and to extend its lower part so that $c(s_{\nu_1})$ is outside the \saferegion of $c(t_{\nu_1}) = c(s_{\nu_2})$, thus guaranteeing property \textbf{P2}. Property {\bf P3} holds by construction.

\noindent \textbf{Case 3.} If none of $\nu_1$ and $\nu_2$ is a $Q$-node, then we combine $\gamma_{\nu_1}$  and  $\gamma_{\nu_2}$ as in Fig.~\ref{fi:snode-3}. We may need to scale one of the two drawings so that their widths are the same. Property {\bf P1} holds, as it holds for  $\gamma_{\nu_1}$  and  $\gamma_{\nu_2}$. Furthermore, we ensure {\bf P2} by performing the following stretching operation. Let $\ell_a$ and $\ell_b$ be two horizontal lines slightly above and slightly below the horizontal segment of $c(t_{\nu_1}) = c(s_{\nu_2})$, respectively. We extend all the vertical segments intersected by $\ell_a$ or $\ell_b$ until the \saferegion of $c(t_{\nu_1}) = c(s_{\nu_2})$ does not intersect any other \saferegion. Property {\bf P3} holds by construction.

Let $\mu$ be a $P$-node of $T$, having $\nu_1$ and $\nu_2$ as children (recall that neither $\nu_1$ nor $\nu_2$ is a $Q$-node, since $G$ is a reduced \ttsp). We combine $\gamma_{\nu_1}$ and $\gamma_{\nu_2}$ as in Fig.~\ref{fi:pnode}. We may need to scale one of the two drawings so that their heights are the same.  Property {\bf P1} holds, as it holds for  $\gamma_{\nu_1}$ and $\gamma_{\nu_2}$. To ensure {\bf P2}, a stretching operation similar to the one described in Case 3 is possibly performed by using a horizontal line slightly above (below) the horizontal segment of $c(s_\mu)$ ($c(t_\mu)$). Property {\bf P3} holds by construction. 

To deal with the time complexity of algorithm \algo, we represent  each cross with the coordinates of its four end-points. To obtain linear time complexity, for each drawing $\gamma_\mu$ of a node $\mu$, we avoid moving all the crosses of its children. Instead, for each child of $\mu$, we only store the offset of the top-left corner of the bounding box of its drawing. Afterwards, we fix the final coordinates of each cross through a top-down visit of $T$. 
The above discussion can be summarized as follows.

\begin{lemma}\label{le:vcct}
Let $G$ be an $n$-vertex reduced \ttsp. Algorithm \algo computes a \balanced and \spaced \UCCT $\gamma$ of $G$ in $O(n)$ time. 
\end{lemma}

\section{ 1-bend Drawings}\label{se:upward}

\begin{figure}[t]
\centering
 \vspace{-20pt}
\subfigure[]{\includegraphics[scale=0.9,page=1]{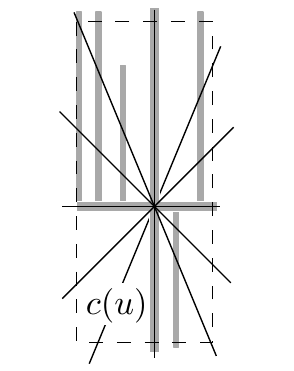}\label{fi:slopes}}\hfil
\subfigure[]{\includegraphics[scale=0.9,page=2]{figures/1bend-upward}\label{fi:1bend}}\hfil
\subfigure[]{\includegraphics[scale=0.9]{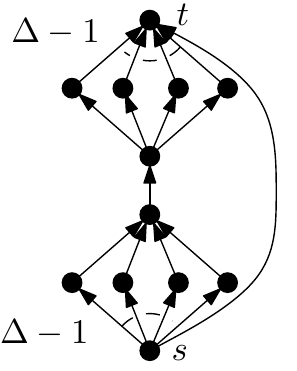}\label{fi:lowerbound}}
\caption{(a)-(b) Transforming an \UCCT into a 1-bend drawing. (c) An \ttsp requiring at least $\Delta$ slopes in any  1-bend upward planar drawing.}
\end{figure}

We start by describing how to transform an \UCCT of a reduced \ttsp into a 1-bend upward planar drawing that uses the \upslopeset $\mathcal S_\Delta$. Let $\gamma$ be an \UCCT of a reduced \ttsp $G$ and let $c(u)$ be the cross representing a vertex $u$ of $G$ in $\gamma$. Let $p_1,\dots,p_{\delta}$ ($\delta \geq 1$) be the contacts along the horizontal segment of $c(u)$, in this order from the leftmost one  ($p_1$) to the rightmost one ($p_\delta$). Let $c$ be either the center of $c(u)$, if $c(u)$ is non-degenerate, or $p_{\lfloor \delta/2 \rfloor+1}$ if $c(u)$ is degenerate. Consider the set of lines $\ell_0, \dots, \ell_{\Delta-1}$, such that $\ell_i$ passes through $c$ and has slope $s_i \in \mathcal S_\Delta$ (for $i=0,\dots,\Delta-1$). These lines, except for $\ell_0$, intersect all the vertical segments forming a contact with the horizontal segment of $c(u)$. If $c(u)$ is not degenerate, then $\ell_0$ coincides with the vertical segment, which has at least one contact. In particular, each quadrant of $c(u)$ contains a number of lines that is at least the number of vertical segments touching $c(u)$ in that quadrant. Since $\gamma$ is \spaced, these intersections are inside the \saferegion of $c(u)$. Hence we can replace each contact of $c(u)$ with two segments having slope in $\mathcal S_\Delta$ as shown in Fig.~\ref{fi:slopes} and~\ref{fi:1bend}. More precisely, each contact $p_i$ of $c(u)$ is replaced with two segments that are both in the quadrant of $c(u)$ that contains the vertical segment defining $p_i$. This guarantees the upwardness of the drawing.  Also, each edge has one bend, since it is represented by a single contact between a horizontal and a vertical segment and we introduce one bend only when dealing with the cross containing the horizontal segment.  
Finally, $\Gamma$ is planar, because there is no crossing in $\gamma$ and each cross is only modified inside its \saferegion which, by the \spaced property, is disjoint by any other \saferegion. Thus, every reduced \ttsp admits a 1-bend upward planar drawing with at most $\Delta$ slopes. To deal with a general \ttsp, we subdivide each transitive edge and compute a drawing of the obtained reduced \ttsp. We then modify this drawing to remove subdivision vertices (technical details can be found in~\cite{}).

Figure~\ref{fi:lowerbound} shows a family of {\ttsp}s such that, for every value of $\Delta$, there exists a graph in this family with maximum vertex-degree $\Delta$ and that requires at least $\Delta$ slopes in any  1-bend upward planar drawing. Namely, if a digraph $G$ has a source (or a sink) of degree $\Delta$, then it requires at least $\Delta-1$ slopes in any upward drawing because each slope, with the only possible exception of the horizontal one, can be used for a single edge. In the digraph of Fig.~\ref{fi:lowerbound} however, the edge $(s,t)$ must be either the leftmost or the rightmost edge of $s$ and $t$ in any upward planar drawing. Therefore, if only $\Delta-1$ slopes are allowed, such edge cannot be drawn planarly and with one bend. Thus, the following theorem holds.

\begin{theorem}\label{th:upwardpsl}
Every $n$-vertex \ttsp $G$ with maximum vertex-degree $\Delta$ admits a 1-bend upward planar drawing $\Gamma$ with at most $\Delta$ slopes and angular resolution at least $\frac{\pi}{\Delta}$. These bounds are worst-case optimal. Also,  $\Gamma$ can be computed in $O(n)$ time.
\end{theorem}

Since every \ttspg can be oriented to an \ttsp (by computing a so-called \emph{bipolar orientation}~\cite{DBLP:journals/dcg/RosenstiehlT86,DBLP:journals/dcg/TamassiaT86}), the next corollary is implied by Theorem~\ref{th:upwardpsl} and improves the upper bound of $\frac{3}{2}(\Delta-1)$~\cite{DBLP:conf/latin/KnauerW16} for the case of {\ttspg}s.

\begin{corollary}\label{co:psn}
The 1-bend planar slope number of {\ttspg}s with maximum vertex-degree $\Delta$ is at most $\Delta$.
\end{corollary}

Our drawing technique can be naturally extended to construct 1-bend planar drawings of two sub-families of biconnected {\ttspg}s using $\lceil \frac{\Delta}{2} \rceil$ slopes. Intuitively, if the drawing does not need to be upward, then for each cross $c(u)$ (see e.g. Fig.~\ref{fi:slopes}), one can use the same slope for two distinct edges incident to $u$. Also, the biconnectivity requirement can be dropped by using one more slope. 

\begin{theorem}\label{th:reduced-flat}
Let $G$ be a $2$-connected \ttspg with maximum vertex-degree $\Delta$ and $n$ vertices. If $G$ is reduced or flat, then $G$ admits a $1$-bend planar drawing $\Gamma$ with at most $\lceil \frac{\Delta}{2} \rceil$ slopes and angular resolution at least $\frac{2\pi}{\Delta}$. Also, $\Gamma$ can be computed in $O(n)$ time.
\end{theorem}


\begin{corollary}\label{co:reduced-flat}
Let $G$ be an \ttspg with maximum vertex-degree $\Delta$ and $n$ vertices. If $G$ is reduced or flat, then $G$ admits a $1$-bend planar drawing $\Gamma$ with at most  $\lceil \frac{\Delta}{2} \rceil +1$ slopes and angular resolution at least $\frac{2\pi}{\Delta+1}$. Also, $\Gamma$ can be computed in $O(n)$ time.
\end{corollary}

\section{Open Problems}\label{se:conclusions}

We proved that the 1-bend upward planar slope number of \ttsp{s} with maximum vertex-degree $\Delta$ is at most $\Delta$ and this is a tight bound. Is the bound of Corollary~\ref{co:psn} also tight? Moreover, can it be extended to any partial $2$-tree?

{\small \bibliography{paper}}
\bibliographystyle{splncs03}

\clearpage

\appendix

\renewcommand\thesection{\Alph{section}}

\section*{Appendix}
\vspace{-5pt}
\section{General \ttsp{s}}\label{ap:di}

\begin{figure}[h]
\vspace{-20pt}
\centering
\subfigure[$G$]{\includegraphics[scale=0.9,page=1]{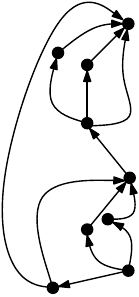}\label{fi:g}}\hfil
\subfigure[$G_r$]{\includegraphics[scale=0.9,page=2]{figures/example}\label{fi:gr}}\hfil
\subfigure[$\gamma$]{\includegraphics[scale=0.9,page=3]{figures/example}\label{fi:uccr}}\hfil
\subfigure[$\Gamma$]{\includegraphics[scale=0.9,page=4]{figures/example}\label{fi:drawing}}
\caption{\label{fi:example}(a) An {\ttsp} $G$; (b) A reduced \ttsp $G_r$ obtained from $G$ by changing the embedding and subdividing the transitive edges of $G$; (c) An \UCCT $\gamma$ of $G_r$; (d) A $1$-bend upward planar drawing $\Gamma$ of $G$ obtained from $\gamma$. }
\vspace{-15pt}
\end{figure}

To deal with a general \ttsp $G$ 
(see, e.g., Fig.~\ref{fi:g})
, we first change the embedding of $G$ as follows. Let $(u,v)$ be a transitive edge, and let $G'$ be the maximal subgraph of $G$ having $u$ and $v$ as poles. We change the embedding of $G'$ so that $(u,v)$ is the rightmost outgoing edge of $u$ and the rightmost incoming edge of $v$. Second, we subdivide $(u,v)$ with a dummy vertex $x$. The resulting graph $G_r$ is a reduced \ttsp 
(see also Fig.~\ref{fi:gr}) 
and therefore we can compute an \UCCT $\gamma$ of $G_r$ 
(see also Fig.~\ref{fi:uccr}), and then turning it into a 1-bend upward planar drawing $\Gamma_r$ of $G_r$, as described above. When doing so, we take care of guaranteeing that the drawings of $(u,x)$ and $(x,v)$ (for each transitive edge $(u,v)$) do not use the horizontal slope (it is not difficult to see that this is always possible). Each transitive edge $(u,v)$ of $G$ is represented in $\Gamma_r$ by a path of two edges $(u,x)$ and $(x,v)$. If at least one between $(u,x)$ and $(x,v)$ is drawn with no bends (i.e., it is drawn vertical), then it is sufficient to remove $x$ to obtain a 1-bend drawing of $(u,v)$. If both $(u,x)$ and $(x,v)$ have one bend, then simply removing the subdivision vertex would lead to a $2$-bend drawing of $(u,v)$. 
\begin{wrapfigure}{r}{0.2\columnwidth}
\vspace{-23pt}
\centering
\includegraphics[scale=0.8,page=2]{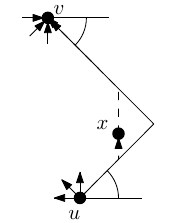}
\caption{\label{fi:transitiveedge}Drawing of a transitive edge.}
\vspace{-20pt}
\end{wrapfigure}
In this case, let $\ell_u$ be the straight line passing through $u$ and the bend of $(u,x)$ and let $\ell_v$ be the straight line passing through $v$ and the bend of $(x,v)$. We obtain a 1-bend drawing of $(u,v)$ by placing a single bend at the intersection point of $\ell_u$ and $\ell_v$ (see also Fig.~\ref{fi:transitiveedge}). Since we did not use the horizontal slope in the drawing of $(u,x)$ and $(x,v)$ such a point exists.  With this operation, the drawing of $(u,v)$ has been extended to the right, and it is possible to modify the construction of the \UCCT $\gamma$ so that $(u,v)$ does not cross any other edge. Namely, when a $P$-node is processed, the algorithm additionally ensures the existence of an empty region where $(u,v)$ can be drawn without crossings. 
For every $P$-node $\mu$, the algorithm ensures that there exists a vertical line $\ell_\mu$ leaving all the contacts of $\gamma_\mu$ on its left and whose horizontal distance from the rightmost side of $R_\mu$ is at least $\frac{h_\mu}{2} \cot \frac{\pi}{2\Delta}$, where $h_\mu$ is the height of $R_\mu$. The region of $R_\mu$ to the right of $\ell_\mu$ is called the \emph{expansion region} of $R_{\mu}$. To achieve the desired width of $R_\mu$, the two (degenerate) crosses $c(s_\mu)$ and $c(t_\mu)$ may be possibly stretched horizontally. Since the expansion region remains empty during the subsequent steps of \algo, edge $(u,v)$ can extend inside this region without creating any crossing. Also, the width of this region is sufficient to contain the $1$-bend drawing of $(u,v)$: the distance between $u$ and $v$ is the height $h_\mu$ of $R_\mu$; the slope of $\ell_u$ is at least $\frac{\pi}{2\Delta}$ and the slope of $\ell_v$ is at most $\pi - \frac{\pi}{2\Delta}$; thus the width of the drawing of $(u,v)$ is at most $\frac{h_\mu}{2} \cot \frac{\pi}{2\Delta}$, which is the width of the expansion region. The resulting drawing $\Gamma$ is a $1$-bend upward planar drawing with at most $\Delta$ slopes (see also Fig.~\ref{fi:drawing}).

\vspace{-5pt}
\section{Undirected Graphs}\label{ap:undi}
\vspace{-5pt}

\begin{figure}[t]
\centering
\subfigure[]{\includegraphics[scale=0.8,page=1]{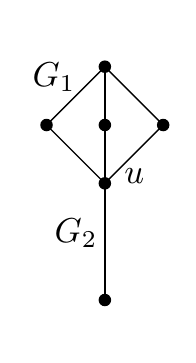}\label{fi:snode-case2-1}}\hfil
\subfigure[]{\includegraphics[scale=0.8,page=2]{figures/snode-case2}\label{fi:snode-case2-2}}\hfil
\subfigure[]{\includegraphics[scale=0.8,page=1]{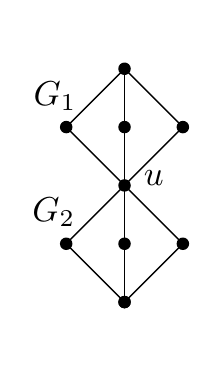}\label{fi:snode-case3-1}}\hfil
\subfigure[]{\includegraphics[scale=0.8,page=2]{figures/snode-case3}\label{fi:snode-case3-2}}\hfil
\subfigure[]{\includegraphics[scale=0.8,page=3]{figures/snode-case3}\label{fi:snode-case3-3}}\hfil
\subfigure[]{\includegraphics[scale=0.8,page=4]{figures/snode-case3}\label{fi:snode-case3-4}}
\caption{Modification of: (a)-(b)  {\bf Case 2} ($\delta_1>1$, and $\delta_2 = 1$); (c)-(d) {\bf Case 3} ($\delta_1>1$, and $\delta_2 > 1$) when both $\delta_1$ and $\delta_2$ are odd; (e)-(f) {\bf Case 3} ($\delta_1>1$, and $\delta_2 > 1$) when both $\delta_1$ and $\delta_2$ are even.}
\vspace{-15pt}
\end{figure}

Consider first a reduced $2$-connected \ttspg $G$ with $n$ vertices, and let $v$ be a vertex of degree two of $G$ (which always exists since \ttspg{s} are partial $2$-trees, and hence 2-degenerate). Let $u$ and $w$ be the two vertices adjacent to $v$, and denote by  $G'$  the graph obtained by removing $v$ from $G$. Graph $G'$ is a connected reduced \ttspg (it may not be $2$-connected anymore). We orient $G'$ to an \ttsp such that $u$ and $w$ are the source and the sink, respectively (this can be done in $O(n)$ time~\cite{DBLP:journals/dcg/RosenstiehlT86,DBLP:journals/dcg/TamassiaT86}). We then compute a \UCCT $\gamma$ of $G'$ by applying the technique of Lemma~\ref{le:vcct}, except for the following modification. 
We aim at guaranteeing that, for each vertex of $G'$ different from $u$ and $w$ that has even degree, either the cross is non-degenerate and both its bottommost and topmost end-points are contacts, or there are two vertically aligned contacts in the middle of the cross (one corresponding to an incoming edge and one to an outgoing edge). In order to achieve this, we need to slightly modify the construction in the case when two graphs are combined in a series composition. Let $u$ be the vertex shared by two \ttspg{s}, $G_1$ and $G_2$, combined in a series composition, and such that the degree $\delta$ of $u$ in $G'$ is even. If $\delta=2$, then $u$ is drawn as a non-degenerate cross touching the two poles of the series composition, and hence we do not need to modify the drawing (see also Fig.~\ref{fi:snode-1}). Suppose that $\delta > 2$, and let $\delta_1$ and $\delta_2$ be the degree of $u$ in $G_1$ and in $G_2$, respectively. When computing the \UCCT of $G_1 \cup G_2$, we apply either {\bf Case 2} or {\bf Case 3} (see also Fig.~\ref{fi:snode-2} and Fig.~\ref{fi:snode-3}) described in Section~\ref{se:ccr}. If we are in {\bf Case 2}, then $\delta_2=1$, and $\delta_1 > 1$ is odd (see, e.g., Fig.~\ref{fi:snode-case2-1}). We then combine the \UCCT{s} $\gamma_1$ of $G_1$ and $\gamma_2$ of $G_2$ such that the middle contact $p_{\lceil \delta_1/2 \rceil}$ of $\gamma_1$ corresponds with the topmost endpoint of the cross representing $u$. In other words, the cross representing $u$ is drawn as a ``T-shape'', as shown in Fig.~\ref{fi:snode-case2-2}. This construction ensures property {\bf P1} for the resulting drawing.  If we are in {\bf Case 3}, then we further distinguish whether $\delta_1$ and $\delta_2$ are both odd or both even. In the first case  (see, e.g., Fig.~\ref{fi:snode-case3-1}), we combine $\gamma_1$ of $G_1$ and $\gamma_2$ of $G_2$ such that the middle contact $p_{\lceil \delta_1/2 \rceil}$ of $\gamma_1$ is vertically aligned with the middle contact $p_{\lceil \delta_2/2 \rceil}$ of $\gamma_2$, as shown in Fig.~\ref{fi:snode-case3-2}.  In the second case  (see, e.g., Fig.~\ref{fi:snode-case3-3}),  we combine $\gamma_1$ of $G_1$ and $\gamma_2$ of $G_2$ such that the  contact $p_{\delta_1/2}$ of $\gamma_1$ is vertically aligned with the contact $p_{\delta_2/2 +1}$ of $\gamma_2$, as shown in Fig.~\ref{fi:snode-case3-4}. In both cases, {\bf P1} is guaranteed.

\begin{figure}[t]
\centering
\subfigure[]{\includegraphics[scale=0.8,page=1]{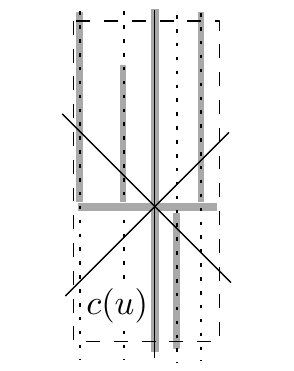}\label{fi:slopes-2}}\hfil
\subfigure[]{\includegraphics[scale=0.8,page=2]{figures/1bend}\label{fi:1bend-2}}\hfil
\subfigure[]{\includegraphics[scale=0.8,page=3]{figures/1bend}\label{fi:deg2}}\hfil
\subfigure[]{\includegraphics[scale=0.8,page=4]{figures/1bend}\label{fi:flat-1}}\hfil
\subfigure[]{\includegraphics[scale=0.8,page=5]{figures/1bend}\label{fi:flat-2}}
\caption{(a)-(b) Transforming an \UCCT into a 1-bend drawing. (c) Reinsertion of vertex $v$. (d) A flat \ttspg $G$, and the maximal subgraph $G'$ having vertices $u$ and $v$ as poles. (e) Graph $G$ with a modified embedding so to ensure that $(u,v)$ will be drawn with no bends.}
\vspace{-15pt}
\end{figure}

Thanks to the described modification, we can turn $\gamma$ into a $1$-bend drawing as follows. Let $c(u)$ be the cross of vertex $u$ in $\gamma$, and let $p_1,\dots,p_{\delta}$ ($\delta \geq 1$) be the contacts along the horizontal segment of $c(u)$, in this order from the leftmost one   to the rightmost one. Let $c$ be either the center of $c(u)$, if $c(u)$ is non-degenerate, or $p_{\lfloor \delta/2 \rfloor+1}$ if $c(u)$ is degenerate. Consider the set of lines $\ell_0, \dots, \ell_{\Delta/2-1}$, such that $\ell_i$ passes through $c$ and has slope $s_i \in \mathcal S_{\lceil \Delta/2 \rceil}$ (for $i=0,\dots,\frac{\Delta}{2}-1$). Differently from the case described in Section~\ref{se:upward}, if $\delta = \Delta$, then each line must be used to draw two contacts of $c(u)$ rather than one. Also, if the number of incoming and outgoing edges of $u$ is different, then these lines may not intersect all the vertical segments forming a contact on the horizontal segment of $c(u)$. Suppose first that $u$ is neither the source nor the sink of the graph, i.e., it has at least one incoming and at least one outgoing edge. Then, our modified construction ensures that the line with vertical slope always intersects two middle contacts of $c(u)$, and thus the above set of lines intersect all the vertical lines supporting the vertical segments that touch $c(u)$. Since $\gamma$ is \spaced, all these intersections are inside the \saferegion of $c(u)$. Hence we can replace each contact of $c(u)$ with two segments having slope in $\mathcal S_{\lceil \Delta/2 \rceil}$ as shown in Fig.~\ref{fi:slopes-2} and~\ref{fi:1bend-2}. Each contact $p_i$ of $c(u)$ is replaced with two segments, which in this case may not be in the same quadrant of $c(u)$. 

Consider now the source $s=u$ and the sink $t=w$ of $G'$. The additional issue for these two vertices is that the vertical slope cannot be used twice, as there are no crosses below $s$ and above $t$. However, since we removed vertex $v$, the degrees of $s$ and $t$ are smaller than $\Delta$, and thus we can avoid to use  the vertical slope twice for these vertices. Thus, we replace the corresponding crosses with two points, and turn all contacts into polylines using at most one bend each. We then reinsert vertex $v$ as follows. We draw a segment from $s$ down to a point $p$ below $s$ using the vertical slope (which is free by construction), and a segment from $t$ up to a point $q$ above $t$ using the vertical slope. We connect $p$ and $q$ with two segments that use a negative and a positive slope of $\mathcal S_{\lceil \Delta/2 \rceil}$, and draw $v$ at their intersection point, as shown in Fig.~\ref{fi:deg2}. Points $p$ and $q$ can be chosen sufficiently far from $s$ and $t$ so to guarantee that no crossing is introduced. The resulting drawing is a $1$-bend planar drawing of $G$ with at most $\lceil \frac{\Delta}{2} \rceil$ slopes.

\medskip

We now turn our attention on flat \ttspg{s}, and show that also for this family of graphs the $1$-bend planar slope number is $\lceil \frac{\Delta}{2} \rceil$.  Let $G$ be a flat \ttspg, and let $T$ be its decomposition tree.  Let $(u,v)$ be a transitive edge of $G$. Let $G'$ be the maximal subgraph of $G$ having $u$ and $v$ as poles and associated with the $P$-node $\mu$ of $T$. By definition of flat \ttspg, the subtree of $T$ rooted at $\mu$ does not contain any $P$-node sharing only one pole with $\mu$. In other words, all the edges incident to both $u$ and $v$ in $G'$ are not in parallel with any other subgraph of $G'$, and therefore $u$ and $v$ have the same degree $\delta$ in $G'$; see also Fig.~\ref{fi:flat-1}. It follows that we can change the embedding of $G'$ such that the edge $(u,v)$ is the $\lfloor \delta/2 \rfloor + 1$-th edge encountered in the counterclockwise circular order of the edges around $v$, starting from the leftmost edge of $v$ (i.e., the edge on the left path of the outer face of $G'$); see also Fig.~\ref{fi:flat-2}. After this operation, we subdivide all transitive edges, and apply the same algorithm described in Section~\ref{se:upward} as modified above to use the slope-set $S_{\lceil \Delta/2 \rceil}$. The constructed embedding guarantees that the two edges incident on each subdivision vertex always use the vertical slope, and thus have no bends. It follows that we can just remove these vertices and obtain a 1-bend planar drawing of $G$ using at most $\lceil \frac{\Delta}{2} \rceil$ slopes, as desired.
The above discussion can be summarized as follows. 

The above discussion can be used to prove Theorem~\ref{th:reduced-flat}. Corollary~\ref{co:reduced-flat} folllows from the fact that the $2$-connectivity requirement of Theorem~\ref{th:reduced-flat} can be dropped if we use at most $\lceil \frac{\Delta}{2} \rceil +1$ slopes. With these many slopes, the vertical slope can be used only once, and thus we do not need to remove a vertex of degree $2$, which requires the input graph to be $2$-connected. 

\end{document}